\documentclass[aps,prl,amsmath,amssymb,superscriptaddress,reprint]{revtex4-1}

\usepackage{graphicx}

\usepackage[space]{grffile} 
\graphicspath{{/Users/cinti/Google Drive/Documents/Bibliography/}} 

\usepackage{epsfig}
\usepackage{bm}
\usepackage{ulem}
\usepackage{color}
\usepackage{soul}
\usepackage[british,UKenglish,USenglish,english,american]{babel}

\begin{document}

\title{Cluster stability driven by quantum fluctuations}

\author{Fabio Cinti}
\affiliation{Department of Physics and Astronomy, University of Florence, Via Sansone 1, I-50019, Sesto Fiorentino (FI), Italy;} \email{fabio.cinti@unifi.it}

\affiliation{Department of Physics, University of Johannesburg, P.O. Box 524, Auckland Park 2006, South Africa.}


\begin{abstract}
By means of an accurate path-integral Monte Carlo we investigate a two-dimensional ensemble of 
particles interacting via a Lifshitz-Petrich-Gaussian potential. In particular, analysing  
structures described by a commensurate ratio between the two wave numbers that 
mark the pattern, the Lifshitz-Petrich-Gaussian boson 
model may display a stable and well-defined stripe phase lacking any global phase coherence but 
featuring a superfluid signal  along the stripe direction only.
Upon increasing quantum fluctuations and quantum-mechanical exchange of bosons,
the double-degeneration of the negative minima in the Fourier transform of the potential 
is removed at the expense of a density modulation peculiar to a cluster triangular crystal. 
We also show that this last structure possess all features adhering to the definition of a supersolid phase. 

\end{abstract}

\pacs{pacs here}

\maketitle

\section{\label{sec1} Introduction}

The physical properties of cluster phases is progressively becoming a pivotal aspect in condensed and 
soft matter as well as in atomic physics. Particle aggregates show properties, whose features can be mainly described microscopically by means of effective potentials and, at equilibrium, they can self-organize in non-trivial structures \cite{SciortinoF.2016,RevModPhys.84.759}. This allows us to design and experimentally control such  structures at different length and energy scales. 
In a pure classical context it is already well established that the 
necessary mathematical condition to aggregate particles into cluster phases is that the Fourier transform  
of the effective two-body potential must exhibit at least a negative minimum \cite{Likos01b,Shin09}. 

In the context of soft matter and biological systems, much work as been already done by successfully  
using generalised exponential models which are capable of accounting the behaviour of 
colloids and polymer chains \cite{Likos2001,PhysRevLett.80.4450,PhysRevE.94.042120}. 
Such models can introduce patterns by balancing between repulsive forces at short range (particles in a cluster) and those at long-range (or intermediate long-range) affecting the rest of the structure. More interestingly,
the pattern symmetry would generate spherical, cylindrical, sheet-like, inverted, or even bicontinuous 
structures ~\cite{Shin09,Glaser07,Fornleitner_2010,Lindquist2016,Sweatman2019}. Moreover, in some cases, the interplay between different length-scales may also produce quasicrystal phases \cite{Dotera2014}. 

Very recently, Barkan \text{et al.}~\cite{Barkan14} has designed a set of isotropic pair potentials which 
are capable of assembling rich wealth of ordered equilibrium structures such as, for instance, stripes or quasicrystals. Interestingly, such two-body potentials furnish a Fourier transform exhibiting an instability at two different wave numbers. If the ratio between the wave numbers is an integer (or more generically a rational number), the potential describes a stripe phase \cite{PhysRevA.15.319,PhysRevLett.79.1261}. Otherwise if the ratio turns out to be irrational one obtains quasicrystals with a 10 and 12-fold symmetry. 

Considering particles obeying quantum statistics, the corresponding many-body physics of quantum 
aggregates is indeed paving the way to new and challenging phases of matters. 
By way of illustration,  alkaline atoms off-resonantly excited to Rydberg states furnish two-body 
soft-shoulder-like shape potentials that may show quantum clusters which result to be identified as  an example of supersolid phase \cite{Pohl2010,Cinti:2014aa,Henkel2012,Labuhn2016,Zeiher2016}. Furthermore,
it has been pointed out that a Bose system interacting via a pair-wise potential composed of a repulsive core at short distances, furnishes quantum-mechanical exchanges that stabilise triangular cluster phase at finite temperature in 
a wider region of parameter space than predicted by calculations in which exchanges are neglected \cite{1367-2630-16-3-033038}. 
Regarding systems made up of fermions, it has been shown in Ref.~\cite{PhysRevB.69.115327}
there exits a competition between quantum-liquid and electron-solid (cluster) phases for some Landau levels
by varying the filling factor.



Quantum aspects of quasicrystals with a dodecagonal symmetry have been newly discussed in Ref.~\cite{1905.12073}.
In particular, the authors observed that a quantum quasicrystal still maintains the dodecagonal pattern 
as well as a small yet finite superfluidity signal. Moreover large quantum fluctuations induce a transition to a triangular cluster and then to a supersolid phase. Such a dynamic may lead to others unforeseen behaviours if quantum effects are taken into account. 
It is worthwhile to mention that the debate concerning the intrinsic properties of quantum stripe patterns still remains open. 
As an example, Masella  \textit{et al.} \cite{PhysRevLett.123.045301} have recently faced the problem on a lattice. They found that the competition between quantum fluctuations and cluster formation may give to 
an anisotropic stripe supersolid phase. 


Here we propose an innovative theoretical investigation considering 
an ensemble of bosons interacting with a Lifshitz-Petrich-Gaussian (LPG) pair potential.
The study pays special attention to commensurate patterns such as the above-mentioned stripe phase. 
We use path-integral Monte-Carlo (PIMC) simulations to show that  stripe phase remains stable if quantum fluctuations are not too large and without supporting any global superfluidity but showing  a phase coherence along stripes only.
Boosting  fluctuations up, we observe, before a complete  melting, an unexpected structural transition 
to a triangular cluster crystal. This last phase is indeed a  supersolid. 


The rest of the paper is organised as follows: In Section II, we introduce the Lifshitz-Petrich-Gaussian pair potential as well as the microscopic model describing the stripes phase in the quantum regime. Section II also outlines the numerical methodology employed as well as the estimators of the thermodynamic observables. In Section III we illustrate our results, whereas conclusions will be reported in Section IV.

\section{\label{sec2} Model, Methodology and Thermodynamic Observables }

\begin{figure}[t!]
\begin{center}
\resizebox{0.8\columnwidth}{!}{\includegraphics{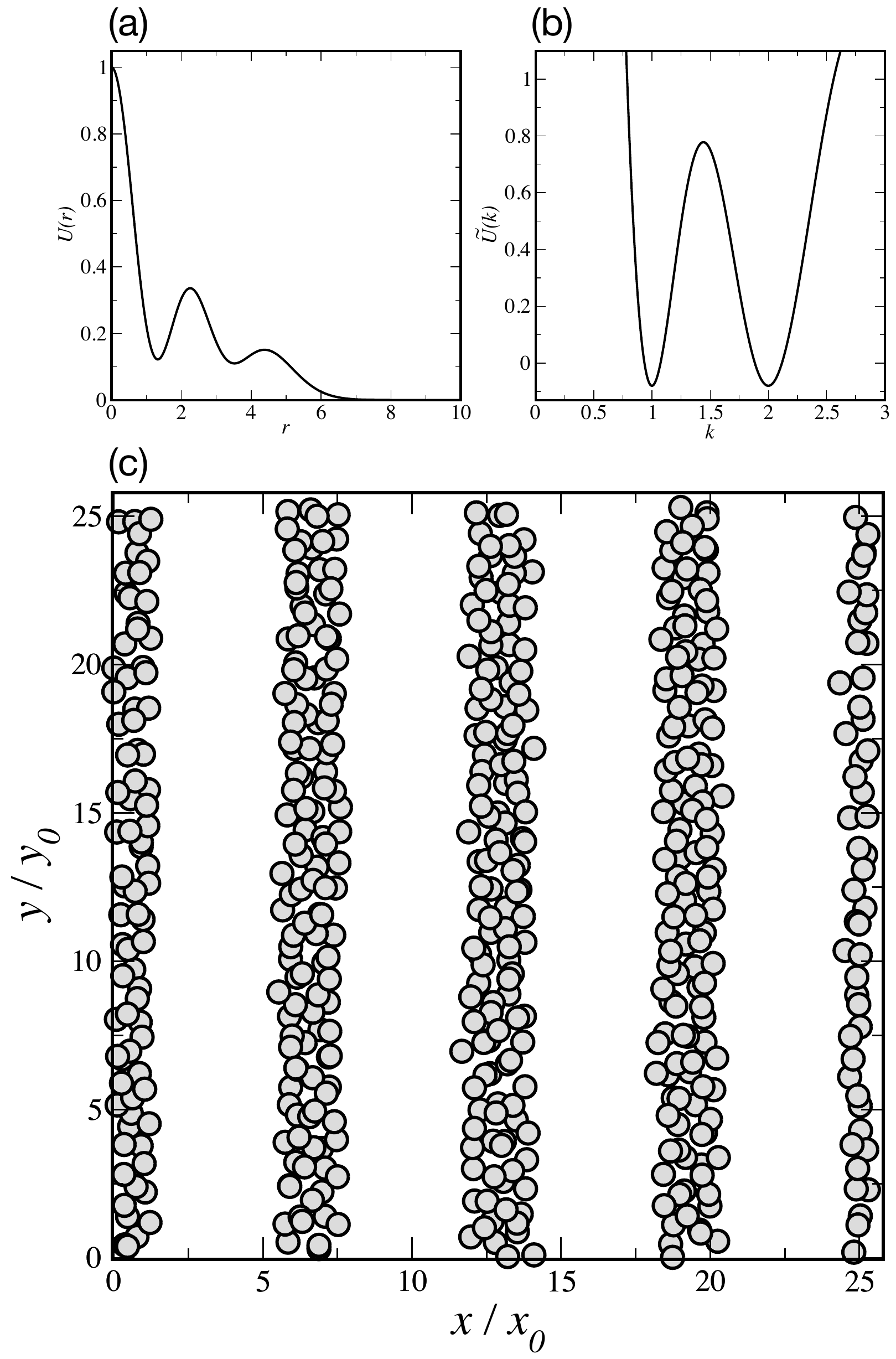}} 
\caption{\label{fig1}  (a) Lifshitz-Petrich-Gaussian pair potential of equation \eqref{eq:pot}
in real space. Parameters $\sigma$ and $C_i$ have been taken from Ref.~\cite{Barkan14}. (b) Fourier transform of  
\eqref{eq:pot}. The panel shows a close view of the minima characterising the classical stripe pattern (c) Snapshot of a stripe phase composed by $classical$ particles interacting  via the two-body potential in (a). The result has been obtained employing a classical Monte Carlo using 512 particles at temperature $t=0.03$.}
\end{center}
\end{figure}

We consider the LPG pair potential~\cite{Barkan14}, defined as 
\begin{equation}\label{eq:pot}
U(r) = e^{-\frac{1}{2}\sigma^2r^2}\left( C_0 + C_2 r^2 + C_4 r^4 + C_6 r^6+ C_8  r^8\right)\,.
\end{equation}
In the present work we pick the parameters $\sigma$ and $C_{i}$ such that the a classical equilibrium configuration at low  temperature forms a striped pattern. Figure~\ref{fig1}a shows the pair potential. 
Fourier transform (Figure~\ref{fig1}b) of \eqref{eq:pot} furnishes two equal-depth negative minima
with a corresponding ratio of commensurate wave-vectors $k=2$~\cite{Barkan14}.
As previously mentioned, we aim to investigate how quantum effects may alter this particular pattern. 
For this purpose, we consider a two-dimensional system composed of $N$ spin-zero bosons  of mass $m$. 
So the Hamiltonian describing the quantum-mechanical system reads 
\begin{equation}
\label{eq:ham}
H=-\frac{\hbar^2}{2m}\sum_{i=1}^{N}\nabla^2_i+\sum_{i<j}^N U\left({\bf r}_{ij}\right)\,,
\end{equation}
where the first term of \eqref{eq:ham} regards the kinetic contribution to the total energy whereas the second sum refers to the 
two-body potential \eqref{eq:pot}, being ${\bf r}_{ij}  = | {\bf r}_i - {\bf r}_j |$
and ${\bf r}_i \equiv (x_i ,y_i )$ the position of $i$-th bosons on the plane, respectively. 

In order to properly quantify the influence of the quantum fluctuations, we introduce
the so-called de Boer parameter~\cite{Kagan2013}
\begin{equation}
\Lambda=\sqrt{\frac{\hbar^2}{mr_0^2 U_0}}
\end{equation} 
where $U_0$ is the pair potential at $r=0$ and $r_0$ is the characteristic length given by the inverse of the wavevector corresponding to the first minimum of the Fourier transform of $U(r)$. In a crystal, the de Boer term accounts for zero-point vibrations 
in the limit for $T\to0$. Depending on the material under investigation, by increasing $\Lambda$ 
the crystal becomes unstable with respect to zero-point motion even at $T = 0$, usually referred as $quantum$ $melting$.
$\Lambda$ is particularly useful to study quantum fluctuation on quantum fluids like He,  Ne and H$_2$ \cite{Sevryuk2010}.

We investigated the equilibrium properties of the system described by the Hamiltonian \eqref{eq:ham} 
employing first-principle computer simulations based on a continuous-space PIMC \cite{Ceperley1995,krauth2006statistical}. 
The calculations include the use of the \textit{worm algorithm} (WA), which allows one to obtain the exact 
thermodynamics properties of a bosonic system. WA has been successfully tested on a large variety of systems, 
including 4He \cite{Boninsegni2006}, Rydberg atoms \cite{Cinti:2014aa,Cinti2010b} and dipolar systems 
\cite{Cinti2015,PhysRevA.95.023622,PhysRevA.92.053625}.
The reader may consult Ref.~\cite{PhysRevLett.96.070601} for a thorough illustration of the methodology.
Here we shall only give a few details concerning the approximation applied on the density operator. 
Given the fact that the LPG  potential in Figure~\ref{fig1}a does not feature 
any dramatic discontinuity, being a smooth analytical function, it is perfectly appropriate to apply a fourth-order expansion 
of the density operators already proposed a while ago by Chin in Ref.~\cite{Chin1997}. The expansion takes into account the first derivatives of the 
inter-particle potential and it has the great advantage of evaluating the density operators with a small
number of time slices, even at low temperature. 

We have worked to find out the equilibrium state of Eq.~(\ref{eq:ham}) at a fixed temperature and number of particles $N$ (canonical ensemble), with $N$ between 256 and 1024. Simulations are performed using periodic boundary conditions along $x$ and $y$-directions. The de Boer parameter is considered in the range between 0 and 1, whereas we set the reduced temperature at $t=k_BT/U_0=0.03$. 

Concerning the system's density, as one can suppose for the tow-body potential here introduced, the kind of pattern that is accomplished at equilibrium is going to result strongly affected by this parameter. Consistently with molecular dynamic simulations discussed in Ref.~\cite{Barkan14}, it is expected to observe a striped phase at a given a reduced density $\rho r_0^2=0.8$. 

In order to test the effective agreement with the molecular dynamic methods, we have first carried out a Monte Carlo simulation  suppressing the first term in Eq.~(\ref{eq:ham}) and analysing the classical limit $\Lambda=0$. The classical simulation 
is then run picking up an initial random configuration which should be regarded as a fluid state. 
Subsequently, the thermodynamic equilibrium at a high enough temperature, $t_0$, is established. 
We recall that the Monte Carlo steps per particle considered here only comprise Metropolis moves at the analysed temperature. $t_0$ has been chosen in order to have a high acceptance ratio per particle \cite{Allen2017}. Then the temperature is decreased gradually $t\to t - \Delta t$ ($\Delta t > 0$) starting with the last configuration sampled at the previous higher temperature. The procedure is completed when the wished temperature is approached. By employing  this painless $annealing$ strategy the stripe pattern can be reached for temperatures $t<0.12$. As an example, Figure~\ref{fig1}c shows the final configuration of the procedure reached down to $t=0.03$ for $N=512$. We note that the chosen $N$ suffices for reproducing a classical configuration made of stable stripes.

\begin{figure}[t!]
\begin{center}
\resizebox{0.8\columnwidth}{!}{\includegraphics{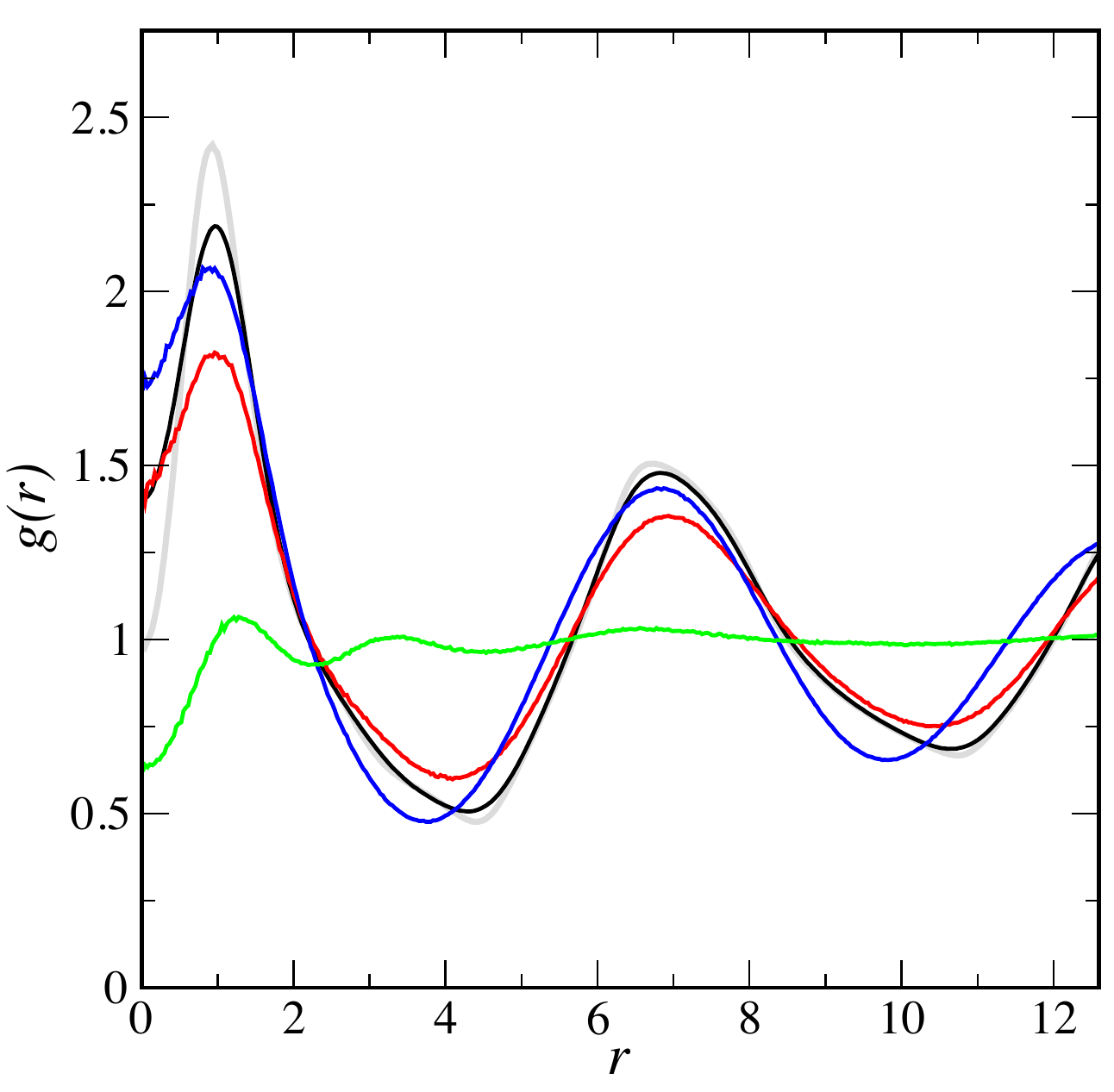}}
\caption{\label{fig2} \textit{Color online}. Radial distribution function $g(r)$ for $\Lambda=0$ (grey line), $\Lambda=0.32$ (black line),
$\Lambda=0.45$ (red line), $\Lambda=0.55$ (blue line) and $\Lambda=0.77$ (green line).}
\end{center}
\end{figure}

Now we move to discuss the thermodynamic observables and their estimators applied in this study. 
Regarding the calculation of the kinetic energy in a PIMC contest, it is well know that 
the evaluation of this estimators can be performed in different ways \cite{Ceperley1995}. In the present work we 
apply the method proposed by Jang \textit{et al.} in Ref.~\cite{Jang2001}. 
This choice is appropriated, again  keeping in mind the smoothness of   
the LPG pair potentials. The lack of discontinuities, in fact, does not spoil the approximation adopted for sampling the density matrix of the system. 

Structural properties of the stripe phase are analysed by means of  the radial distribution function $g(r)$, 
which in the PIMC formalism reads 
\begin{equation}\label{eqgr}
g(r) = \frac{1}{2 \pi \rho r_0^2 (N-1) r} \bigl\langle\sum_{i\,,j\neq i}  \delta \bigl(r-r_{ij}(\tau)\bigr) \bigr\rangle_\tau\,,
\end{equation}
where and $\langle\ldots\rangle_\tau$ is the average over complex time, $\tau$, 
trajectories ${\bf r}_i(\tau)$ \cite{Ceperley1995}. In a generic modulated pattern or a triangular crystal phase Eq.~\eqref{eqgr} displays pronounced and wide maximum and, at the same time, well-defined minima. 
In contrast, a standard superfluid shows a $g(r)$ having the usual liquid-like behaviour, 
that is, a first peak around the averaged inter-particle distance followed by a series of damped oscillations 
around one \cite{chaikin2000principles}. 

\begin{figure*}[t!]
\begin{center}
\resizebox{0.9\textwidth}{!}{\includegraphics{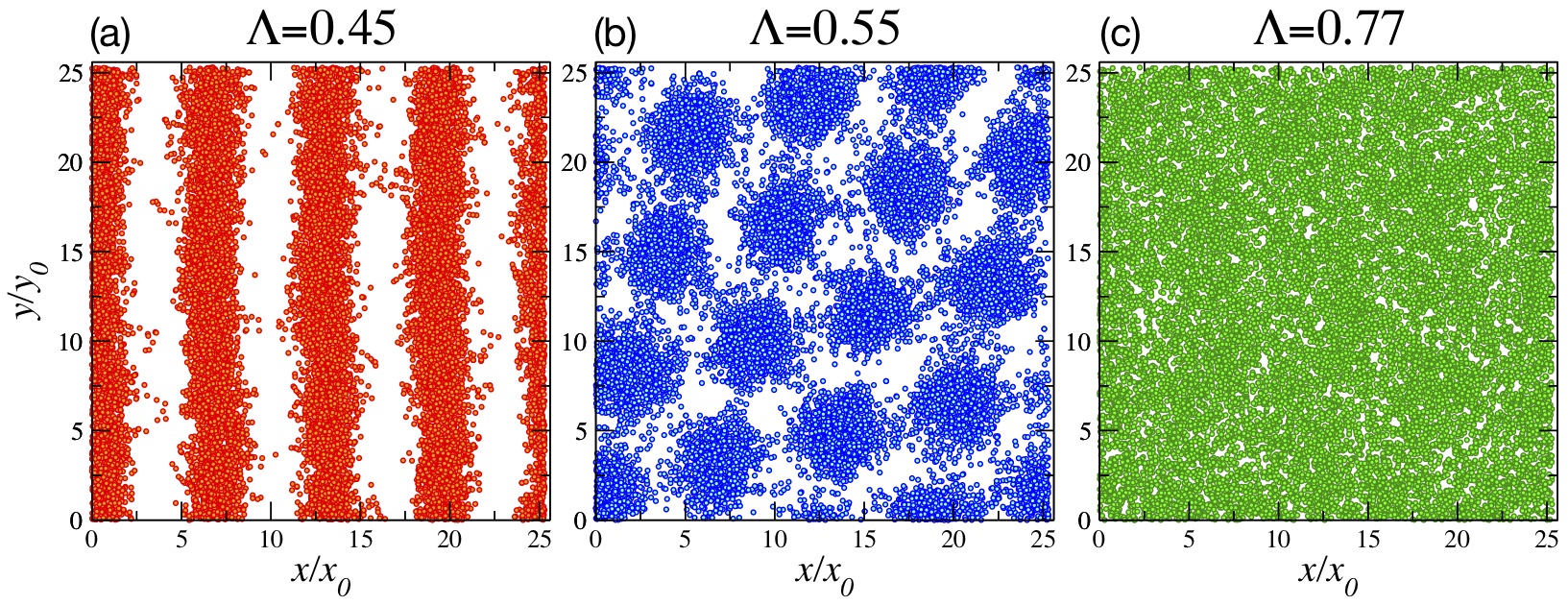}  }
\caption{\label{fig3} \textit{Color online}. PIMC's density distribution in real space of three snapshot configurations increasing $\Lambda$: (a) stripe phase, (b) supersolid phase and (c) superfluid phase (see text).}
\end{center}
\end{figure*}

A further observable extremely useful in order to address quantum properties of Eq.\eqref{eq:ham}
is the superfluid fraction of the system. Specifically, within the two-fluid model, the superfluidity is obtained 
taking advantage of the linear response theory by inspecting the different 
response to boundary motion of the normal component respect to the superfluid one.
Here we implement the estimator proposed a few decades ago by Pollock and Ceperley \cite{poll87}. 
Following this, the superfluid fraction as a function of temperature, along the orthogonal directions $x$ and $y$ of the simulated box, yields
\begin{equation}\label{superformula}
f_{s}^{(i)}=\frac{t}{\Lambda^2\,\rho\,r_0^2}\,\langle w_{i}^2\rangle,
\end{equation}
where $i=\,x,\,y\,$, and $\langle \ldots \rangle$ stands for the  thermal average of the 
winding number estimator $w_i$ \cite{poll87}. Differently from an homogeneous superfluid, 
a density-modulated superfluid (supersolid) presents a non-unitary but uniform 
response of the estimator in Eq.~\eqref{superformula}  \cite{Cinti:2014aa}.

\section{\label{sec3}Results}

\begin{figure}[b]
\begin{center}
\resizebox{0.8\columnwidth}{!}{\includegraphics{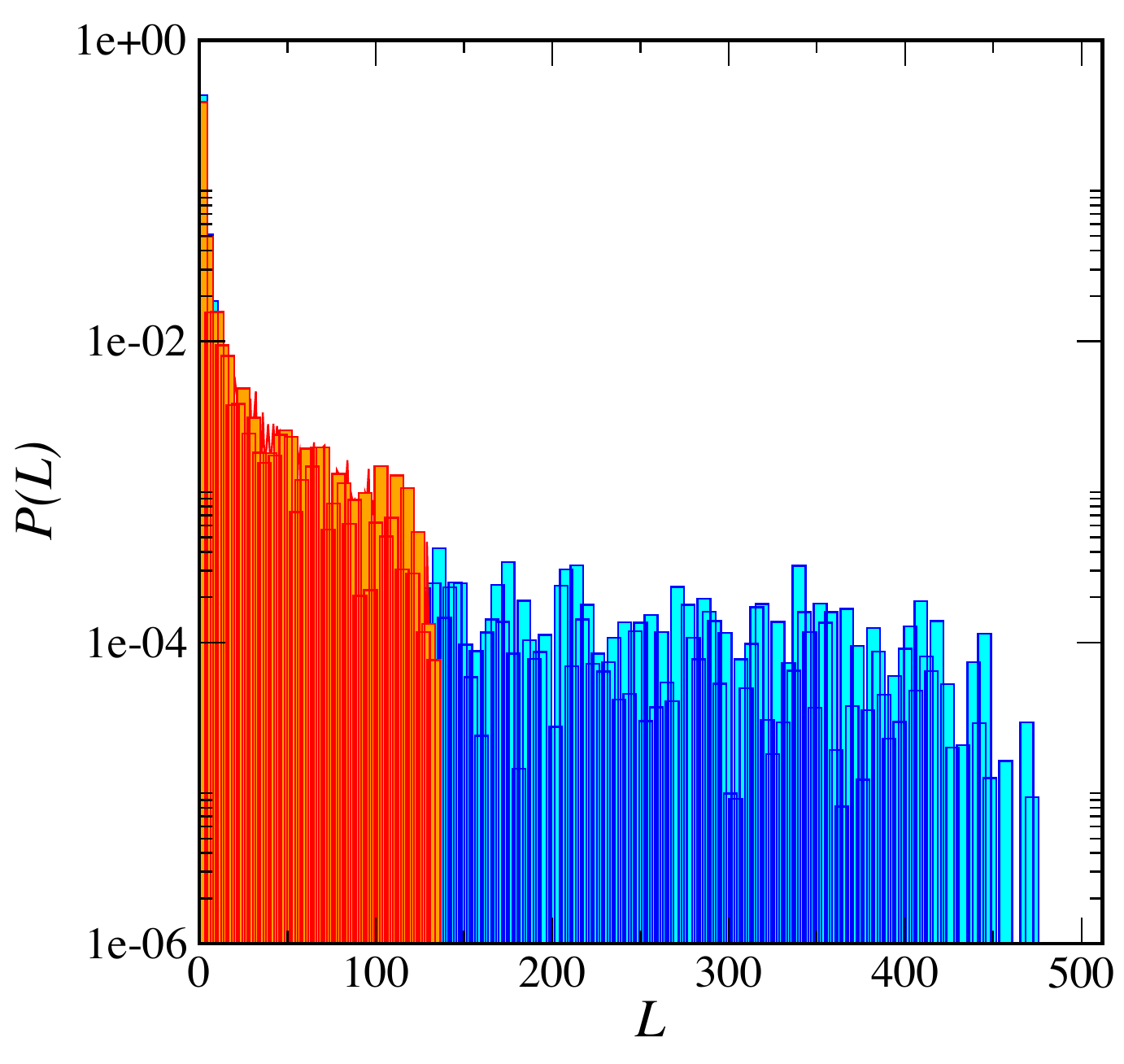}}
\caption{\label{fig4} \textit{Color online}. Frequency of exchange cycles of length $L$ ($1\leq L\leq N$) in the stripe phase (red histogram), in which the superfluidity results finite along the stripe direction only, and  supersolid phase (blue histogram), in which the superfluidity is uniformly finite throughout the simulated box.}
\end{center}
\end{figure}

We begin with examining the stability of the stripe patterns increasing $\Lambda$. Figure~\ref{fig2} displays the radial distribution function considering  some different values of the de Boer parameter. The case for $\Lambda=0$ (grey line) reproduces the density modulation for the classical system, previously discussed (Figure~\ref{fig1}c). Considering the quantum regime ($\Lambda\ne0$), one can roughly discern two different behaviours: one referring to modulated phases ($\Lambda=0.32,\,0.45,\,0.55$) and a second ($\Lambda=0.77$ ) identifying a superfluid phase.  For $\Lambda=0.32$ and 0.45 (black and red line, respectively) the stripe phase mainly shows the same modulation of the classical configuration. Yet at $r=0$, $g(r)$ for $\Lambda=0.32,\,0.45,\,$ and 0.55 turns out to be larger than its classical counterpart. The effect outlines the increasing of the local fluctuations when the Bose--Einstein statistics is taken into consideration. Interestingly, the behaviour at $r=0$ grows stronger about $\Lambda=0.55$. At the same time, again at $\Lambda=0.55$,  we also obtain a change of density modulation thus marking a cluster solid (the corresponding snapshot configuration is offered in Figure~\ref{fig3}b). In addition, exhibiting a finite superfluid signal, this crystal can be regarded as a supersolid (see below). Finally for $\Lambda\gtrsim 0.6$ the supersolid evidently gets into a superfluid. 

\begin{figure}
\begin{center}
\resizebox{0.8\columnwidth}{!}{\includegraphics{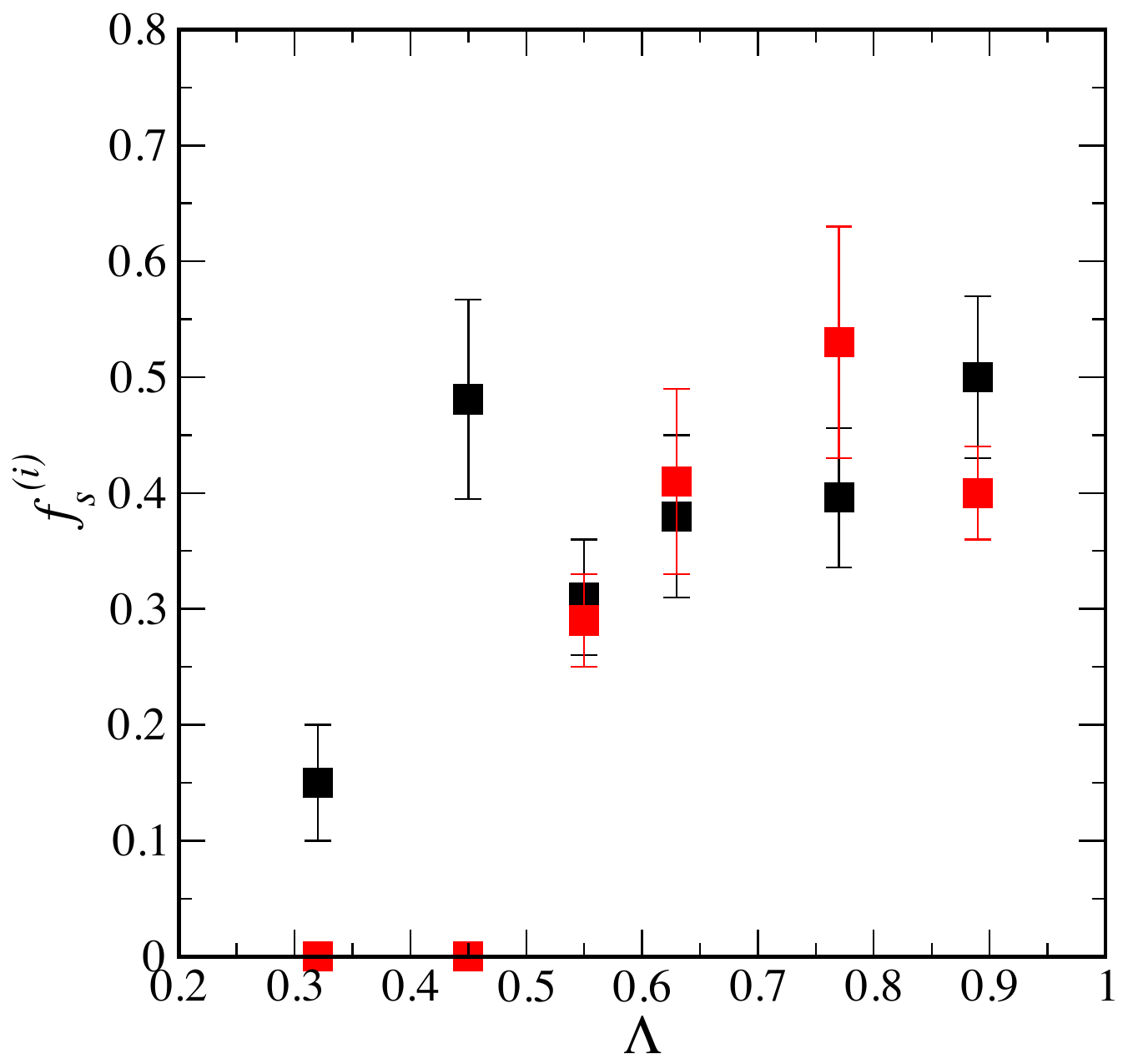}}
\caption{\label{fig5} \textit{Color online}. 
Superfluid fraction $f_s^{(i)}$, $i=x,y$, as a function of the de Boer parameter along the stripes direction (in this work $f_s^{(y)}$, black square) and orthogonally ($f_s^{(x)}$, red square) to them.}
\end{center}
\end{figure}

Figure~\ref{fig3} shows snapshots of the projection of world lines onto the $xy$-plane for three different phases, once again modifying the de Boer parameter. Here the system displays a well defined structural transition about $\Lambda\approx 0.5$, between stripe phase (panel a) and a triangular cluster supersolid (panel b). Bosons in Figure~\ref{fig3}a visibly delocalise themselves along a single stripe and not among nearest neighbour stripes. This initial analysis looks to exclude the existence of a global coherence, although for properly addressing the issue we have to discuss the estimator of the superfluid fraction (see Figure~\ref{fig5}). As indicated above, panel b shows clear evidence that quantum fluctuations lead to the transition from stripes to supersolid. In panel c we show a configuration made of bosons completely delocalised throughout the box, identify then a superfluid phase ($\Lambda=0.77$). 

A comparison between configurations in Figure~\ref{fig3}a and b can be also qualitatively operated taking into consideration the frequency of cycles of permutations among bosons  \cite{citeulike:13652852,Jain2011}. As one expects, long exchanges of identical particles take place in a system where a non-zero global superfluid response is observed. Not only long exchanges characterise a usual homogeneous superfluid (of which, for instance, Figure~\ref{fig3}c is a peculiar example) but also regimes where a spatial broken symmetry may endure, marking then the presence of a supersolid phase \cite{RevModPhys.84.759}. The probability of permutation $P(L)$ involving $L$-bosons (with $1\leq L\leq N$) is reported in Figure~\ref{fig4}. $P(L)$ referring to a stripe phase (red histogram) shows that permutations entail cycles to about 130 bosons, that is approximatively the number of particles located on each stripe. On the contrary, in the supersolid regime (blue histogram), we observe permutations extending themselves over long cycles and almost covering the entire set of particles in the box ($L\lesssim N$).

Figure~\ref{fig5} depicts the superfluid fraction -- Eq.~\eqref{superformula} -- by varying the de Boer parameter. In the stripe phase ($\Lambda\lesssim0.5$),  $f_{s}^{(i)}$ displays a strong anisotropy. In particular, the superfluid fraction is finite along the $y$-direction and it vanishes along the $x$-direction ($f_{s}^{(x)}=0$ and $f_{s}^{(y)}\ne0$, respectively). This result conclusively asserts that each stripe is phase coherent, but globally the system is not; or, more precisely, the system behaves like a collection of independent $quasi$-superfluid chains. It is interesting to mention that the lack of global coherence (i.e. meaning the absences of supersolidity) has been also recently noted on systems of striped dipolar bosons \cite{Cinti2019}.

For $\Lambda\gtrsim 0.5$ the superfluid signal results finite and uniform along both directions ($f_{s}^{(x)}\approx f_{s}^{(y)}$). Considering the fact that we are working at finite temperature,  superfluidity can not help to plainly discern between cluster supersolid and superfluid phase. On the other hand such a demarcation can be easily operated checking up quantities like permutation cycles or equilibrium configurations. Yet, one would tentatively state that $f_{s}$ at $t=0.03$ is increasing upon increasing $\Lambda$ from supersolid to superfluid phase. By all means, superfluidity appears to remain unaffected for $\Lambda\gtrsim0.65$, that is, when bosons are regarded as completely delocalised throughout the box, signalling then the complete melting of any quantum modulated phase. 


\begin{figure}
\begin{center}
\resizebox{0.8\columnwidth}{!}{\includegraphics{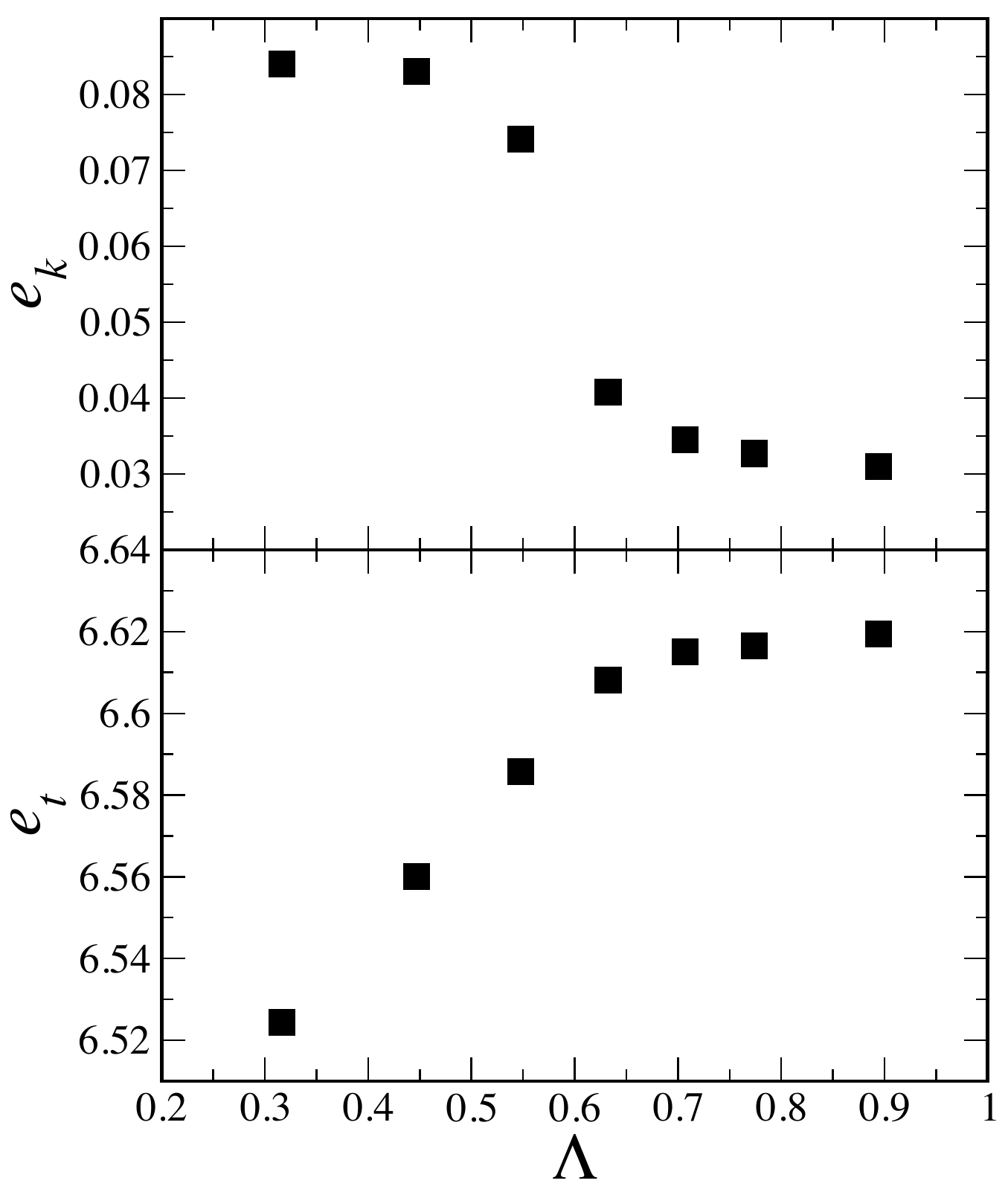}}
\caption{\label{fig6} Kinetic energy (upper panel) and total energy (lower panel) as a function of $\Lambda$. 
Error bars lie within point size.}
\end{center}
\end{figure}

Now we examine the stability of the phases discussed so far introducing some considerations connected to the system's energy. Figure~\ref{fig6} reports the kinetic energy per particle ($e_k$ upper panel) as well as the total energy per particle ($e_t$ lower panel) again versus the de Boer parameter at $t=0.03$. Total energy shows two different slopes that are related, the first to the ordered phases and the second to the superfluid phase. Regarding  the kinetic energy, as one might expect, we observe that $e_k$ decreases when quantum fluctuations get stronger and stronger. The transition between the cluster crystal and a fluid phase is marked by a clear jump about  $\Lambda\approx0.6$, around this value the $e_k$ is halved.  At lower $\Lambda$ the kinetic energy displays an additional lowering around the transition between stripe and cluster phase thus also here it sounds reasonable to reaffirm that  the effect is mainly driven by quantum fluctuation and quantum mechanical exchanges.

%

\section{\label{sec4}Discussion and conclusions}

In this work we have investigated the stability of a stripe phase at finite temperature and introducing quantum fluctuations into the system.
The stripe pattern formation is classically introduced by considering a two-body LPG potential \cite{Barkan14}. 
For moderate value of the de Boer parameter ($\Lambda\lesssim0.5$) the stripe phase still survives against 
the tendency of bosons to delocalise throughout the box. The pattern furnishes a superfluid signal 
along the stripe direction only, without evidence of any global coherence. The preset results is consistent
if compared to a recent study on a two-dimensional system of bosons interacting via a dipole-dipole interaction \cite{Cinti2019}. 
In particular, authors observe crystal stripe phases if dipoles are tilted beyond a certain critical value. 
Also for dipolar bosons, the global phase coherence does not emerge and consequently without showing 
any supersolid phase. Interestingly, the analogy between Ref.~\cite{Cinti2019} and the stripe phase introduced by the LPG potential are indeed quite evident. Considering their quantum features 
both systems behave as a set of one-dimensional uncorrelated boson Luttinger liquids. 

The LPG boson model also shows that quantum fluctuations drive a structural transition from stripe 
to a cluster triangular crystal, apparently where thermal fluctuation does not play any specific role. 
Indeed, in the range $0.5\lesssim\Lambda\lesssim0.6$, we remark that $(i)$ 
the competition between quantum fluctuations and the LPG potential tend to remove the degeneration of the Fourier transform of Eq.~\eqref{eq:pot} and 
to impose a density modulation with a wavelength corresponding to a single equilibrium minima; 
$(ii)$ quantum-mechanical exchanges of bosons particles act to stabilize the solid phase, as already observed in Ref~\cite{1367-2630-16-3-033038}.
By increasing $\Lambda$ further, the competition between the two term get lost and cluster 
crystals melts into a superfluid. 

To conclude, extending the present study, future work may address how similar complex structures can control quantum-mechanical exchanges, considering other two or multi-lengthscale soft-core potentials, 
possibly applicable in an experimental contest as, for instance, ultra-cold dipolar atoms 
\cite{Bottcher2019,PhysRevA.96.053630,Tanzi2019,Chomaz2019,PhysRevA.96.013627,PhysRevLett.119.215302,Kora2019,PhysRevLett.123.015301}
or supporting  the understanding of other engaging systems such as quantum quasicrystals \cite{PhysRevLett.111.185304,PhysRevLett.120.060407,Viebahn2019}.

\textit{Acknowledgements.} 
The author wishes to thank P.~Ziherl and T.~Macr\`i for valuable discussions. 

\bibliographystyle{apsrev4-1} 

%

\end{document}